\documentclass[12pt,a4paper]{article}
\usepackage{graphicx}
\usepackage{natbib}
\usepackage{latexsym}
\usepackage{bm}

\bibpunct{(}{)}{;}{a}{,}{,}

\title{Egocentric Path Integration Models and their Application to
  Desert Arthropods}

\begin{document}

\maketitle

\begin{center}
Tobias Merkle, Martin Rost and Wolfgang Alt
\end{center}

\begin{center}
Theoretical Biology,\\
Mathematical and Natural Science Faculty,\\ 
University of Bonn\\
Kirschallee~1, D-53115~Bonn, Germany
\end{center}

\begin{abstract}
Path integration enables desert arthropods to find back to their nest
on the shortest track from any position. To perform path integration
successfully, speeds and turning angles along the preceding outbound
path have to be measured continuously and combined to determine an
internal {\em global vector} leading back home at any time. A number
of experiments have given an idea how arthropods might use allothetic
or idiothetic signals to perceive their orientation and moving
speed. We systematically review the four possible model descriptions
of mathematically precise path integration, whereby we favour and
elaborate the hitherto not used variant of egocentric cartesian
coordinates. Its simple and intuitive structure is demonstrated in
comparison to the other models. Measuring two speeds, the forward
moving speed and the angular turning rate, and implementing them into
a linear system of differential equations provides the necessary
information during outbound route, reorientation process and return
path. In addition, we propose several possible types of systematic
errors that can cause deviations from the correct homeward
course. Deviations have been observed for several species of desert
arthropods in different experiments, but their origin is still under
debate. Using our egocentric path integration model we propose simple
error indices depending on path geometry that will allow future
experiments to rule out or corroborate certain error types. 
\\
\end{abstract}

\newpage
\section{Introduction}
\subsection{Path integration}

Desert arthropods display the ability to return from a foraging
excursion back to the nest on a straight way, often called the {\em
  home vector}. This ability to make a bee-line to the nest (or
another location, such as a feeding site) without orientation on
visible markers is based on an internal mechanism of {\em path
  integration} or {\em dead reckoning}, i.e.\ an integration of
walking speed and angular variation along the arthropod's walking
route. The result is a {\em global vector} that enables the arthropod
to determine distance and direction of its nest at any position and
time. After detecting and loading up the food, the arthropod just
unreels that vector and, therefore, stays on the right and shortest
track to its nest. Charles Darwin was the first to assume that animals
may navigate this way \citep{dar_73}. About a century later first
detailed investigations concerning arthropods
\citep{jan_57,goe_58,goe_66,weh_68,jan_70} were conducted. During the
last 30 years, the general interest has focused more and more on the
return path to the nest after foraging, and many investigations on both
arthropods \citep[e.g.][]{weh_86,mul_88,BIS99,col_99} and mammals
\citep[e.g.][]{mit_80,seg_93,ben_96,ben_97,seg_98} have been
performed.

Apart from path integration, it has been shown that many species are
capable of using landmarks to get their bearings
\citep{hof_85a,col_98,bis_03a,weh_03a}. These landmarks, often
referred to as {\em local vectors}, even seem to be preferred to the
global vectors \citep{weh_96a,col_98} as navigational tool. However,
before these vectors can be applied successfully, some information
about their position has to be stored by the arthropod. Moreover,
orientation with the aid of local vectors is error-prone, since
landmarks can disappear or change their appearance, and, of course, is
out of question for nests with no visible landmarks nearby.

The global vector gets updated on the complete trip, even if the
orientation is conducted by using landmarks
\citep[e.g.][]{weh_96a,col_98,col_03a}. Moreover, it has been
demonstrated that, after a sudden failure of the stored landmarks,
desert ants \emph{Cataglyphis fortis} revert to their global vector
for orientation. Even if not used for several days, desert ants keep
the global vector stored in their memory \citep{Zie_97}. Thus, the
relevance of global vectors as the main toolkit for homing seems to be
as clear as the evolutionary necessity to develop abilities to measure
the angular and linear components of the movements and to integrate
them for having a home vector available whensoever.

A number of models \citep{jan_57,mit_73,mul_88,ben_90,GAL90} try to
point out mechanisms of path integration \citep[for reviews,
  see][]{ben_95,mau_95,bie_00}. Whereas earlier models use {\em geocentric}
coordinates to represent the current global vector and the anticipated
home direction, the more recent models are based upon the assumption
that internal calculation of the global vector, pointing from the
animal's head to an anticipated nest position, should be performed in
an {\em egocentric} coordinate system. In this case, however,
mathematical computations and error estimations have so far been
carried out only in \emph{polar coordinates}, describing the
distance and direction towards the nest with respect to the animals
current position and walking direction.

Here we introduce a consistent model for {\em egocentric} path
integration using \emph{cartesian coordinates}. The mathematical
computation of current nest position relative to the moving arthropod
reduces to an inhomogeneous linear differential equation system of two
variables, namely \emph{forward moving speed} and \emph{angular
  turning rate}. Different ways of their estimation and further
processing implemented into this model provide several possibilities
to reproduce some observed reorientation phenomena of arthropods.

The article is organised as follows. We finish the introductory
Section reviewing more details on signal and information processing in
homing arthropods. The next Section 2 is devoted to previous models of
path integration. Section 3 presents our model approach and compares
it to those in the previous Section. In Section 4 we turn to
systematic errors in homing paths which have been studied extensively
and may give hints about internal information processing. Taking
advantage of the simple structure of our model we propose error tests
for future experiments. Finally we discuss our results and briefly
comment on possible neural realizations of egocentric cartesian path
integration and possible generalisations including global vectors for
feeding sites.

\subsection{Required information for path integration}  

Allothetic and idiothetic signals provide the arthropod with the
information required for path integration. During the following
description of these signals, we shall focus on desert arthropods and
here, in particular, on desert ants \emph{Cataglyphis fortis} and
\emph{Cataglyphis bicolor}, because more is known on them as compared
to other arthropods. Nevertheless, we shall also mention similar
investigations on other arthropods in order to firmly base the
modelling principles, mainly because neurobiological analyses have been
performed on larger and more easily accessible arthropods 
\citep{weh_03a}.

\subsubsection{Allothetic signals}

Without doubt, the main allothetic signals to be considered are visual
inputs. Among them are landmarks which, however, have limited overall
value as described above. More reliable turn out to be optical sources
indefinitely far away: with regard to desert ants, spectral skylight
gradient, sun position, and the pattern of polarised skylight are the
most important cues
\citep{WEH97a,WEH97b,weh_01a,weh_03a,WEH03b}. These three can work
without help of each other, as was shown in experiments where one or
even two of them had been made inoperative \citep{mul_88,WEH97a}. In
all three the ants continuously use the sky as a reference to determine
their body axis orientation.

Whenever the arthropod applies {\em spectral cues}, it makes use of
the fact that light waves with their different wavelengths are not
equally distributed over the illuminated sky. This ability was shown
by \cite{WEH97a} for ants and \cite{ros_84} for bees.

Direct orientation with respect to the {\em azimuthal position} of the
sun (or any other light source) has been found in ants and many other
arthropods (e.g.\ bees, \citealp[von][]{fri_50}; or spiders,
\citealp{goe_58}).

The {\em skylight polarisation} pattern, also referred to as skylight 
compass, represents the most effective and stable means for
orientation in desert ants. Desert ants are able to see the e--vector
polarisation pattern that is produced by scattering of the sunlight at
air molecules in the atmosphere. This ability has been found in other
arthropods and vertebrates as well, first of all in bees
\citep[von][]{fri_49}, but also, for instance, in desert locusts
\emph{Schistocerca gregaria} \citep{EGG93}, desert isopods
\emph{Hemilepistus reaumuri} \citep{HOF84} or desert beetles 
\emph{Parastizopus armaticeps} \citep{BIS99}.
A small visible section of the sky has been shown to be sufficient to
detect rotations \citep{fen_86,weh_94,WEH97b}.

Each rotation of the ant's body axis results in a corresponding change
of this orientational angle allowing the ant to measure not only its
current body direction relative to the allothetic skylight pattern,
but also the rate of its angular rotation, independent of whether it
is moving or turning on spot.

Although the polarised light pattern is changing with elevation of the
sun, desert ants are able to use a stereotypical projection that
resembles the skylight pattern at dawn or dusk, respectively, in their
memory \citep{WEH97a,WEH97b,weh_98,weh_01a}. Unlike for the
task to find a feeder at different daytimes (\citealp{WEH87};
\citealp{weh_93a}; \citealp{dye_94}) sun movement need not be
compensated to find back during an excursion, which normally lasts
only a few minutes, so the resulting error is negligible.

Although \cite{ron_95} have shown that frontal optic flow influences
the ants' odometer, the mechanisms to use allothetic cues for
detecting {\em directions} and {\em rotations} seem to be
inappropriate for measuring {\em distances} or {\em speeds}.
Therefore, additional tools making use of idiothetic signals are needed.

\subsubsection{Idiothetic signals}

Far less is known about the use of idiothetic signals for the
orientation of desert arthropods. Compared to allothetic signals they
seem to be of minor or no importance for detecting
directions. Experiments with desert ants have shown that it is
possible to predict navigational errors by manipulating the visible
section of the sky \citep{WEH97a,WEH97b,weh_98,weh_01a}. Hence, the
ants obviously do not even revert to proprioceptive signals if the
polarisation compass as standard tool provides strange results. Also
desert beetles \emph{Parastizopus armaticeps} rely on the position of
the light source and the polarisation compass and seem not to revert
to proprioceptive cues in the case of ambiguities
\citep{BIS99}. Desert ants \emph{Cataglyphis fortis}, when captured at
a feeder and transferred to a test area in a dark flask without any
allothetic signals available, immediately after their release do
reorientate and set out into their stored home direction (T.~Merkle,
personal observation). Thus, any possibly existing idiothetic signals
do not have an effect on the ant's reorientation under such
conditions.

On the other hand, \cite{ron_95} found that desert
ants are able to estimate their walked distances without allothetic
signals. Therefore they proposed that ants use odometers that mainly
rely on proprioceptive signals. This is backed by investigations that
could eliminate energy consumption as possible cue for measuring
speeds or walking distances, when ants walk along slopes \citep{woh_02} or
with heavy load \citep{sch_93}. It seems quite obvious that such
proprioceptive signals derive from movements of the legs (for bristles
as mechanoreceptors cf.\ \citealp{kei_97}). 

\section{Previous models of path integration}
\label{prevmod}

\subsection{Geocentric models}
\label{prev_geocent}

The term path integration was established by \cite{mit_73}, referring
to a simple and evident mathematical algorithm, namely to determine
the ant's current position  by integrating its moving direction
$\bm{\theta}(s)$ along the migration path or, equivalently, by
integrating its velocity vector $\bm{V}(t) = v(t) \; \bm{\theta}(t)$ over
time $t$, where $v(t)$ denotes the ant's forward speed.
This yields the estimated final positional vector $\bm{P}$ from start
to end point of the path. The estimated global {\em home vector} is
then the inverse vector $\bm{G} = -\bm{P}$.

\subsubsection{Cartesian coordinates}
\label{prev_geo_cart}

\cite{mit_73,MIT82} use \emph{cartesian coordinates} to represent the
integrated positional vector $\bm{P}=(x,y)$ and the current
directional vector $\bm{\theta} = (\cos \phi, \sin \phi)$, where the
animal's angular orientation $\phi$ is given relative to some
reference direction, e.g.\ skylight polarisation. This is chosen as
the initial moving direction of the arthropod in their case study of
the spider \emph{Agelena labyrinthica}. In their model the arthropod
is supposed to estimate the current angle $\phi$ in two different
ways: by using an \emph{idiothetic azimuth storage} as integrated
value of its proprioceptively measured turning rates $\omega(t)$ along
the previous path (this idea of azimuth integration is related to the
earlier theory by \citealp{jan_57}), or by directly measuring the
\emph{allothetic azimuth value} $\phi$ of the current body axis with respect
to an exogenous direction. 
Then, the resulting two inputs of the directional vector 
$\bm{\theta}$ are weighted and summed up for path
integration. With the aid of this model the authors could reproduce
typical two--segment experiments, during which a light source had been
turned by $90^\circ$.

It should be noticed that the described model of \cite{mit_73} is a
closed-loop control system (\emph{``Wirkungsgef\"uge''}) applicable to
any excursion of an arthropod, including the outbound route towards a
food source position $\bm{P}_0$, whose cartesian coordinates have to
be internally stored, or the homing route towards the origin
$\bm{P}_0=(0,0)$. In both cases the control system produces an
efferent motor signal for the turning angle $\omega$ being negatively
proportional to $|\bm{P}_0 - \bm{P}| \sin \delta$, where $\delta$ is
the deviation angle between current moving direction $\phi$ and the
global vector $\bm{G} = \bm{P}_0 - \bm{P}$.  By this steering
algorithm, the arthropod will turn into the direction of the global
vector and walk towards $\bm{P}_0$ until the global vector is
zero. Thus, the path integration mechanism is supposed to work during
the whole excursion.

This comprehensive navigation model has been adapted to experiments
with rodents \citep[e.g.][]{mit_80,ben_97} or humans \citep{mit_01}
and recently up--dated to be consistent with results on neural
activity patterns in the hippocampus of mammals \citep{mit_00}. It
provides important principles of information processing, path
integration, and motor control, and is mathematically easily
realizable in computer simulation programs.
Nevertheless, it remains
open, whether and how the necessary computational steps of calculating
trigonometric functions are physiologically performed within the
neural nets of arthropods or mammals. Moreover, this model supposes
two successive integrating mechanisms including the necessity to store
the computed variables, first $\phi = \int \omega \, dt$ and then
$\bm{P} = \int (\cos \phi, \sin \phi) \, ds$, if integration is 
over walked distance, $s$ denoting arc length, or, 
$\bm{P} = \int (v \bm{\theta})(t) \, dt$, if integration is over time. 
In the last case, besides determination
of the turning speed $\omega$, also that of forward speed $v$ is
required (see Table 1).

The extensive work on neural ``head cell'' and ``place cell'' dynamics
in mammals could be mentioned here, with the remarkable property that
a change in angular information, represented by head cells, can induce
a corresponding rotation of the two--dimensional activity pattern in
the imaginary chart represented by the array of place cells
\citep{sam_97,mit_00}. With regard to {\em desert ants}, \cite{weh_03a}
postulates the existence of a certain number of ``compass neurons'',
each with an accurately defined compass direction resulting in a
maximum firing rate of the respective neuron, whenever the arthropod
is heading into that direction. 

\subsubsection{Polar coordinates}
\label{prev_geo_pol}

Geocentric models in {\em polar coordinates} have been developed and
applied during the last 20 years by \cite{weh_86} and \cite{mul_88},
taken up by \cite{har_95} in connection with dynamical representations
by cyclical neural chains. In order to represent the actual position
vector $\bm{P} = r \, (\cos\nu, \sin\nu)$ of the arthropod, the
proposed model algorithms require the animal to compute, at least
approximately, distance $r$ from the nest and angle $\nu$ of the
position vector relative to an allothetic reference direction, 
determined by the sun or, most frequently, by polarised skylight.
Then, the global home vector is $\bm{G} = -\bm{P} = r \,
(\cos(\nu+\pi), \sin(\nu+\pi))$.

For a segmented path with step length $s_n$ (taken to be $1$ for
simplicity), \cite{mul_88} derive approximate recursive formulas for
updating the polar coordinates $(r_n,\nu_n)$ after the $n$th moving
step. The only additional input needed during each step, besides
knowing step length (or measuring forward speed), is the angle
$\tilde \delta_n = \phi_n - \nu_n$ between moving direction $\phi_n$
and the direction $\nu_n$  of the positional vector. In a continuous
description the corresponding general path integration formulas read
$r = \int \cos(\phi-\nu) \; ds$ and $\nu = \int \sin(\phi-\nu)/r \;
ds$  and are equivalent to a
system of nonlinear ordinary differential equations (see also Table 1)
\begin{eqnarray}
\frac{dr}{dt} & = & v \; \cos(\phi-\nu) \label{e01} \\
\frac{d\nu}{dt} & = & \frac{v}{r} \; \sin(\phi-\nu). \label{e02}
\end{eqnarray}
Again, as in Section 2.1.1, this mathematical integration algorithm 
requires the ability to
calculate the nonlinear trigonometric functions and, in addition, to
perform the division by distance $r$. \cite{mul_88} and \cite{har_95}
suggested the trigonometric functions could be approximated by
piecewise linear or polynomial functions. This led to a systematic
misestimation of increments in both variables, $r$ and $\nu$, for
moving directions not parallel or antiparallel to the position vector,
i.e.\ for $\tilde \delta = \phi - \nu \neq \pm \pi$. \cite{mul_88}
thus could remarkably well reproduce systematic errors in the angular
component $\nu$  of the global vector, observed in the classical
two--segment experiments, not only for desert ants but also for most
arthropods and mammals \citep[e.g.][]{bis_57,goe_58,mul_88,seg_93}. In
Section 4.3 we give a detailed analysis of this topic.

\subsection{Egocentric models}
\label{prev_egocent}

Another approach, which appears to be more adequate but came into
consideration much later, is to model the path integration process of
a moving animal in terms of a moving coordinate frame centred around
the animal's body, thus reflecting the fact that it perceives all
sensory inputs relative to its own position and
orientation. \cite{ben_90} chose \emph{polar coordinates} to represent
the global vector $\bm{G} = -\bm{P} = r \, (\cos \delta, \sin
\delta)$, where now the reference direction for $\delta = 0$ is the
body axis, serving as $X$--axis of the corresponding  cartesian
coordinate frame with the orthogonal lateral $Y$--axis, see
Fig. 1. Although the distance variable $r$ is the same as in the
geocentric polar model, the derived recurrent formulas for updating
$r_n$ and $\delta_n$ turn out to be much more complicated than any
other formula used before. First, egocentric cartesian coordinates
$(X_n,Y_n)$ are updated in terms of the former polar ones,
\begin{eqnarray}
X_{n+1} & = & r_n \, \cos(\delta_n - \omega_n) - s_n \label{e03} \\
Y_{n+1} & = & r_n \, \sin(\delta_n - \omega_n) \label{e04}
\end{eqnarray}
where $\omega_n$ denotes the change of the direction and $s_n$ the
length of the subsequent step. (An interchange of the order of stepping
and turning would give $s_{n+1}$ in (\ref{e03}) but no fundamental
change). Then, these equations are transformed into the new egocentric
polar coordinates:
\begin{eqnarray}
	r_{n+1} & = & \sqrt{{X_{n+1}}^2 + {Y_{n+1}}^2} \label{e05}\\
	\delta_{n+1} & = & \arctan \left( {\frac{Y_{n+1}}{X_{n+1}}}
	\right) \label{e06}
\end{eqnarray}
(in order to calculate the correct values of $\delta_{n+1}$ the signs
of $X_{n+1}$ and $Y_{n+1}$ have to be considered, see
\citealp{ben_95}).

The advantage of this egocentric model is that now the only
\emph{input variables} are step length $s_n$ and turning angle
$\omega_n$, or, in the corresponding continuous path integration
model, forward speed $v$ and angular turning rate
$\omega$. \cite{GAL90} considered the corresponding differential
equation using egocentric polar coordinates in the continuous limit of
infinitely small time steps which, in the corrected formulation by
\cite{ben_95}, are
\begin{eqnarray}
\frac{dr}{dt} & = & - v \; \cos \delta  \label{e07} \\
\frac{d\delta}{dt} & = & v \; \frac{\sin \delta}{r} -
\omega. \label{e08}
\end{eqnarray}
These equations can directly be obtained form the corresponding polar
coordinate equations,  Eqs.~(\ref{e01}) and (\ref{e02}), by performing 
a simple angular transformation, $\delta = \nu + \pi - \phi$, so that 
again they require to compute division by
$r$ and trigonometric functions (see also Table 1).

In their simulation analysis, \cite{ben_90} studied the influence of
random errors on the estimation of changes of direction and walking
distance. With regard to directional changes, they distinguished
between allothetical and idiothetical orientation: they considered
idiothetical estimation as ``measuring the change of direction
itself'', whereas the allothetical estimation is defined as ``a
comparison between the heading of current and previous step relative
to some exteroceptive compass'' \citep{mau_95}. The different
estimation procedures were realized by providing the actual values
with normal distributed errors.

In their simulations random errors of allothetic signals had only
little influence, whereas those of idiothetic signals lead to
noticeable misestimation. \cite{ben_95} conclude that the egocentric
coding process is quite sensitive to idiothetic errors and organisms
relying on allothetic cues for measuring directional changes by far
outmatch those relying on idiothetic cues.

\section{Cartesian model for egocentric path integration}
\label{sectcartmod}

\subsection{System of linear differential equations for the global vector}

\begin{figure}[!ht]
\begin{center}
\includegraphics[width = 10cm]{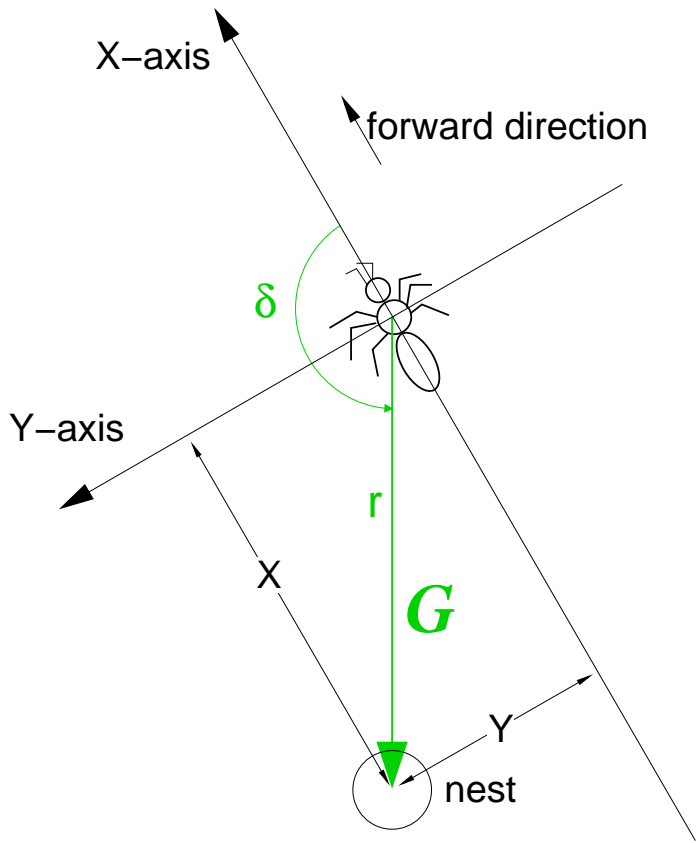}
\end{center}
\caption{Theoretical scheme of egocentric path integration by means of
  cartesian coordinates $X$, $Y$ specifying the position of the nest
  relative to the arthropod's body axes and determining the global
  vector $\bm{G} = (X,Y)$, here with $X < 0$, $Y > 0$. In contrast,
  the corresponding model in polar coordinates (Section
  \ref{prev_egocent}) uses the two variables $r$, distance to the nest,
  and  $\delta$, angle between head orientation and nest direction.}
\label{fig01}
\end{figure}

The physiological sensing and locomotion apparatus of any arthropod is
completely bound to its body. It is therefore naturally related to its
two symmetry axes, the posterior-anterior axis and the perpendicular
right-left axis. Thus, when identifying these symmetry axes with the
$X$ and $Y$ axis of a cartesian coordinate frame $(X,Y)$ and taking
the arthropod's body centre as the origin $(0,0)$, this
constitutes a proper planar moving coordinate frame for representing
the {\em relative position} of any object in the planar neighbourhood
of the arthropod, e.g.\ its nest. In this {\em egocentric cartesian}
model the global vector pointing from the arthropod's body to the
nest, relative to the animal's actual body axis orientation, is just
$\bm{G} = (X,Y)$, corresponding to the same vector as in Section
\ref{prev_egocent}, there only written in polar coordinates, see
Fig.~\ref{fig01}.  Notice that the original ansatz by \cite{ben_90}
already mentioned this cartesian coordinate system, but then switched
to polar coordinates for path integration (see Section
\ref{prev_egocent}). Indeed, the continuous version of 
Eqs.~(\ref{e03}) and (\ref{e04}), given the  arthropod's
{\em forward speed v} and {\em angular turning rate $\omega$}, 
yields the following model equations for a precise update of 
the global vector $(X,Y)$ during motion
\begin{eqnarray}
	\frac{d X}{d t} & = & - v + \omega \cdot Y \label{e09}	\\
	\frac{d Y}{d t} & = & - \, \omega \cdot X \label{e10}
\end{eqnarray}   
Compared to all other continuum equations or analogous discrete
recursion algorithms developed previously (see Section 2), this
two--dimensional differential equation system is remarkably simple: It
is linear in the two variable quantities $X$ and $Y$, and it just uses
the two speed input parameters as additive or multiplicative terms,
$v$ representing the rate of shifting the $X$ coordinate backwards,
$\omega$ the rate of rotating the $(X,Y)$ frame clockwise. These
operations can be easily performed by any suitable elementary analogue
circuit network like the one shown below in Fig.~\ref{fig03}. Possible
physiologically realizable neural representations are briefly
addressed in the Discussion Section~\ref{discussion}.

In conclusion, this {\em cartesian egocentric path integration model},
considered as a precise `dead reckoning' system, offers the
most simple computational scheme to determine the global vector and,
simultaneously, being related to a coordinate frame of the moving
arthropod. For comparison with the other models see
Table~\ref{table1}. 

Clearly, biological solutions of difficult problems can be complex and
the simplicity of our model does not make its realisation more likely
than that of other models. From a conceptional point of view the
existence of a simple solution is nevertheless striking, and in the
following we shall present its implementation and results.

\begin{table}[!h]
\hspace{-2cm}
\begin{tabular}[t]{ll|cl|l}
& & Input variables & Internal variables & Global vector \\
\hline
\vspace{-3mm}
& & & & \\
\parbox{2cm}{Geocentric}
& 
\parbox{2cm}{Cartesian \\ {\it (Sect.~\ref{prev_geo_cart})} \\ Polar
  \\ {\it (Sect.~\ref{prev_geo_pol})} }
&
\parbox{2.5cm}{$\omega$,$v$ or $\omega_n$,$s_n$ \\ \\ $\phi$ \\ $s_n$ or $v$}
&
\parbox{3.5cm}{$\phi$, $\bm{P} = (x,y)$ \\
  $\langle$lin./nonlin.\ ODE$\rangle$ \\ $r$ and  $\nu$ \\
  $\langle$nonlinear ODE$\rangle$}
&
\parbox{4cm}{$\bm{G} = \bm{P}_0 - \bm{P}$ \\ \\ $\bm{G} = - r \; (\cos
  \nu,\sin \nu)$}
\\
\vspace{-3mm}
& & & & \\
\hline
\vspace{-3mm}
& & & & \\
\parbox{2cm}{Egocentric}
& 
\parbox{2cm}{Polar \\ {\it (Sect.~\ref{prev_egocent})} \\ Cartesian
  \\ {\it (Sect.~\ref{sectcartmod})} }
&
\parbox{2.5cm}{$\omega$,$v$ or $\omega_n$,$s_n$}
&
\parbox{3.5cm}{$r$ and $\delta$ \\ $\langle$nonlinear ODE$\rangle$ \\
  $X$ and  $Y$ \\ $\langle$linear ODE$\rangle$}
&
\parbox{4cm}{$\bm{G} = r \; (\cos \delta , \sin \delta )$ \\ \\
  $\bm{G} = (X,Y)$}
\\
\hline
\end{tabular}
\caption{Path integration models with the parameters and variables
used for input, internal calculation and output as global vector. For
notations and more details see the various model descriptions in the
text.
}
\label{table1}
\end{table}

\subsection{Modelling foraging excursions, reorientation, and homing}
\label{modelexcurs}

Our path integration model
depends on the values of $v$ and $\omega$, cf.~Table 1. 
These can be regarded as the elementary physiological control
variables which the arthropod uses to steer its locomotion, e.g.\ by
changing speed or frequency of striking leg motion on both sides or,
respectively, on one side relative to the other.

In order to model typical paths of directionally persistent random
walks, as observed for desert arthropods, one has to account for mean
values and standard deviations of speed $v$ and turning rate $\omega$
as well as for their temporal auto-correlations which can be extracted
from corresponding experimental time series
\citep[see][]{ALT90}. Discrete correlated random walk models sometimes
used \citep[e.g.][]{bye_01} are not adequate as they assume piecewise
constant walking directions $\phi_n$ and turning angles
$\omega_n$. The two speeds $v(t)$ and $\omega(t) = d\phi(t)/dt$,
however, being related to the physiologically controlled, relatively
fast leg movement on both sides of the arthropod, should better be
modelled as fluctuating continuous processes on an adequate smoothness
level. The simplest stochastic process of this kind is described by
the following two independent Ornstein--Uhlenbeck equations for first
order coloured noise, which have previously been used also for
modelling the systematic search of arthropods \citep{alt_95},
\begin{eqnarray}
	d v & = & \frac{1}{T_v} \; (v_0 - v)  \; dt + \beta_v \; dW_t
	\label{e11} \\
	d \omega & = & \frac{1}{T_\omega} \; (\omega_0 - \omega) \; dt
	+ \beta_\omega \; dW_t \label{e12}	 
\end{eqnarray}
Following the standard approach, random perturbations are expressed as
additive Wiener increments $dW_t$ \citep{ITO65}. In simulations one
uses a sequence of values $v$ and $\omega$ with {\em finite} time
differences $\tau$,
\begin{eqnarray}
  v_{t+\tau} & = & v_t + \frac{\tau}{T_v} \; (v_0 - v_t) +
  \beta_v \sqrt{\tau} \; \zeta \\
  \omega_{t+\tau} & = & \omega_t + \frac{\tau}{T_\omega} \; (\omega_0
  - \omega_t) + \beta_\omega \sqrt{\tau} \; \zeta
\end{eqnarray}
where $\zeta$ denotes a standard normally (${\cal N}(0,1)$) distributed
random variable, drawn independently at each step for each
variable. Eqs.~(\ref{e11}) and (\ref{e12}) are obtained in the limit
$\tau \to 0$. $v_0$ and $\omega_0$ are the preferred values of forward
and turning speed, respectively, and the $T_{v|\omega}$ denote the
corresponding mean persistence times of fluctuations with amplitudes
$\beta_{v|\omega}$. In case of a stationary time series they yield
variances of size $\sigma_{v|\omega}^2 = \beta_{v|\omega}^2
T_{v|\omega}/2$ by an equilibrium of perturbations with strength
$\beta_{v|\omega}$ and decay at rate $1/T_{v|\omega}$. In our
presentation we assume, for simplicity, that $T_v$ is negligibly
small, such that during locomotion the forward moving speed has a
constant value $v \equiv v_0$. The presented results also hold for the
general case of fluctuating forward speed.

\begin{figure}[!ht]
\begin{center}
\includegraphics[width = 10cm]{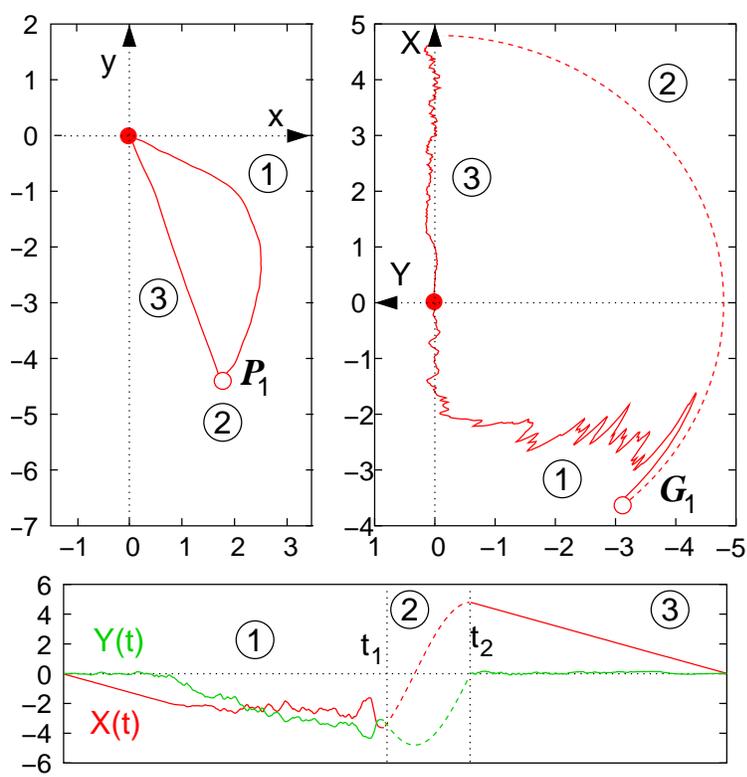}
\end{center}
\caption{Model simulation of a natural outbound path and the
  successful return path back to the nest due to precise path
  integration, according to the 3-Phase-model, see text. Parameters
  used for calculations are, (1) for the outbound path $T_\omega =
  0.3$~s, $\beta_\omega = 1$~s$^{-3/2}$, and constant forward speed $v
  \equiv v_0 = 0.2$~m/s, (2) constant $\omega_{\rm rot} = 1$~s$^{-1}$
  during rotation and (3) for homing the same as in (1) but a feedback
  constant $c = 1/0.05$~s$^{-1}$ for beacon steering. (Top left): Plot
  of the actual path in cartesian $(x,y)$-coordinates of an
  observer. Position of the nest at $(0,0)$ is marked by a filled
  circle. (Top right): Corresponding plot of the nest position in the
  same scale in relative cartesian
  $(X,Y)$-coordinates, where the origin denoting the home position is
  marked by a filled circle. Note that the animal's head direction is
  the $X$--axis pointing upwards, while the lateral $Y$--axis points
  to the left. (Bottom): Corresponding plots of $X$ and $Y$ over time.
}
\label{fig02}
\end{figure}

%
%

In the following, we describe the three successive phases of an arthropod's
typical excursion, using the egocentric path integration system
Eqs.~(\ref{e09}), (\ref{e10}) and the physiological motion control system
Eqs.~(\ref{e11}) and (\ref{e12}). We explain the corresponding
dynamics of the global vector by means of a simulated example
presented in Fig.~\ref{fig02}.

\begin{figure}[!ht]
\begin{center}
\includegraphics*[width = 14cm]{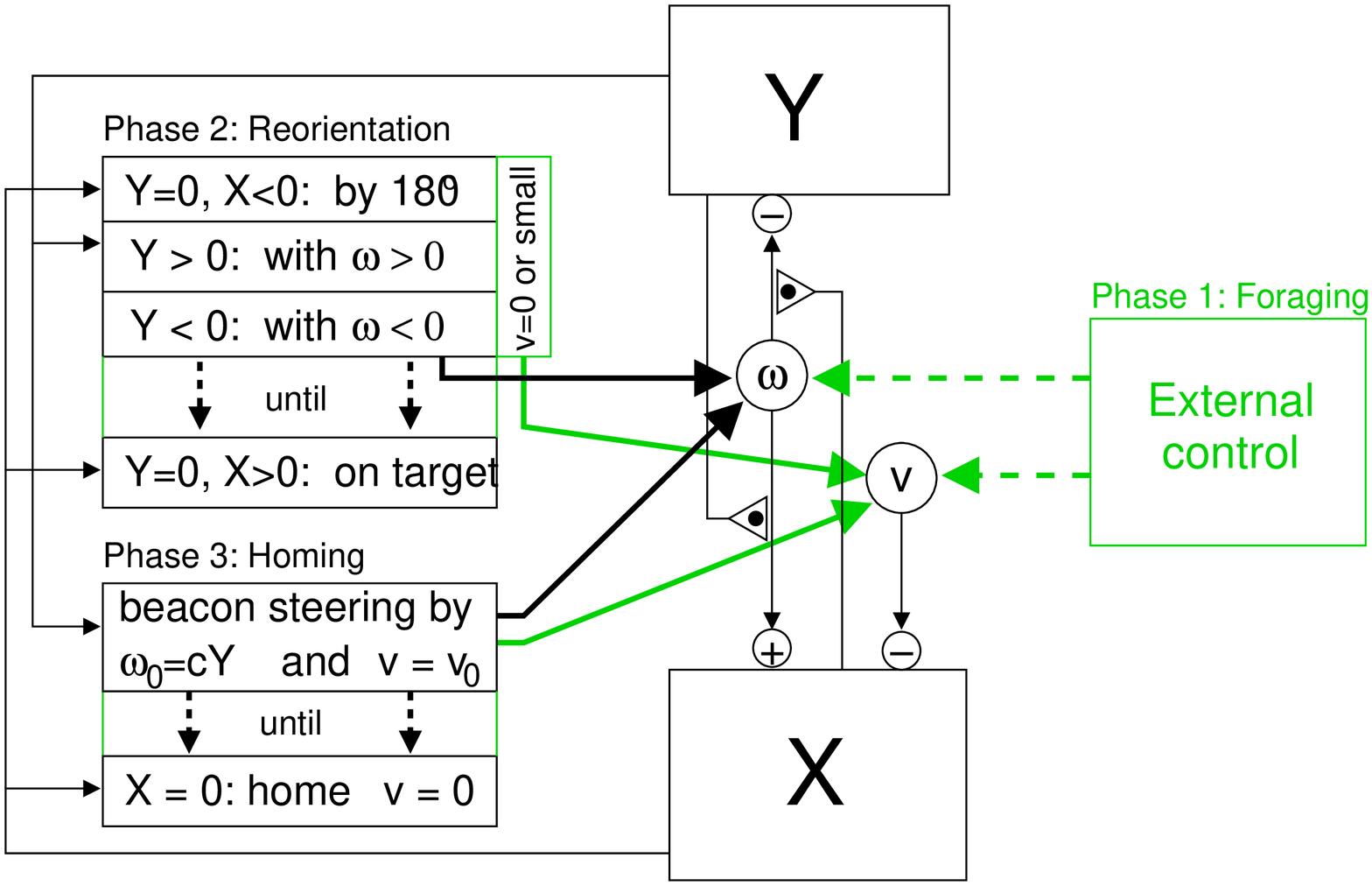}
\end{center}
\caption{Analogue circuit scheme of the egocentric cartesian path
  integration model: dynamics of the two variables $X$ and $Y$
  according to the differential equations~(\ref{e09}) and (\ref{e10})
  and its coupling to the physiological control parameters
  represented by the two speed parameters for turning, $\omega$, and
  forward locomotion, $v$. In {\bf Phase 1}, $\omega$ and $v$ are
  externally controlled (random search, trained path towards feeder,
  \dots). During {\bf Phase 2} and {\bf Phase 3} the $X$ and $Y$
  values feed back into the speed control conditions such that, by
  counter--steering with respect to the `internal beacon' $Y=0$ in the
  latter case, equations~(\ref{e09}), (\ref{e10}) and (\ref{e12})
  constitute a coupled nonlinear control system along the homing
  path.}
\label{fig03}
\end{figure}

\begin{itemize}
   \item{Phase 1:} {\bf (Foraging)} \\
     The arthropod starts foraging at the nest site, e.g.\ $(x_0,y_0)
     = (0,0)$, where the global vector is reset to zero $\bm{G} = (X,Y)
     = 0$. Holding the mean turning rate $\omega_0 = 0$, the animal
     approximately keeps its chosen initial direction, $\phi =
     \phi_0$, for some time, leading it almost straight away from the
     nest, corresponding to increasingly negative $X$ values of the
     internal global vector, while the $Y$ component stay close to
     zero. This initial behaviour is well expressed in the example of
     Fig.~\ref{fig02}, then followed by a random right-hand turn of
     the $(x,y)$-path, which corresponds to increasingly $Y$ values
     meaning that now the nest lies to the right side the animal.
   \item{Phase 2:} {\bf (Reorientation)} \\
     After finding food at some position $\bm{P}_1 = (x_1,y_1)$, the
     arthropod stops there in its current angular orientation, $\phi =
     \phi_1$, keeping the actual global vector $\bm{G}_1 = (X_1,Y_1)$
     internally stored (even during handling the food). Then the
     arthropod starts its reorientation phase by turning on spot, say
     with constant rotation speed $\omega \equiv \pm \omega_{\rm
     rot}$, depending on whether the stored global vector $\bm{G}_1$
     has positive or negative $Y_1$ value. During rotation the global
     vector $\bm{G}=(X,Y)$ also rotates according to path integration
     in Eqs.~(\ref{e09}) and (\ref{e10}), since now we set $v \equiv
     v_0=0$. Finally, the arthropod is assumed to stop its rotation
     ($\omega =0$) as soon as the condition $Y=0$ is fulfilled,
     meaning that now its head is oriented towards the nest and the
     actually positive $X$-value represents the arthropod's distance
     from the nest (cf. the scheme in Fig.~\ref{fig03}). For the
     example in Fig.~\ref{fig02}, see the counter-clockwise rotation
     circle ending on the positive $X$ axis. Notice that the condition
     ${Y=0}$ corresponds to ${\delta = 0}$ in egocentric polar
     coordinates (Section 2.2 and Fig.~\ref{fig01}) because of the
     equivalences  $Y = r \cdot \sin\delta$ ($\approx r \delta$ for
     small $\delta$) and $X = r \cdot \cos\delta$ ($\approx r +
     r\delta^2/2$ for small $\delta$).
   \item{Phase 3:} {\bf (Homing)} \\        
     The arthropod now returns back to the nest (as straight as
     possible) according to the actually stored global vector
     $\bm{G}=(X,0)$. Since during walking the global vector will be
     constantly updated and, due to inevitable random perturbations,
     the $Y$ component will eventually deviate from the zero value,
     the arthropod must tend to hold the internal {\em steering
     condition} $Y=0$ as closely as possible. This can be modelled by
     implementing a {\em counter-steering} turning rate $\omega_0 = -
     c Y$ into the stochastic differential equation (\ref{e12}). See
     Fig.~\ref{fig03} for an analogue circuit scheme describing this
     feedback control, which nonlinearly and cyclically couples the
     linear path integration system, Eq.~(\ref{e09}) and (\ref{e10}),
     to the linear motor control equation (\ref{e12}).

     Finally, in our model the arthropod is assumed to stop its return
     phase as soon as the $X$ value of its global vector becomes
     zero. The resulting home path in the simulated example of
     Fig.~\ref{fig02} clearly shows, how some random perturbations
     lead to small deviations in the homing direction of the $(x,y)$
     path and corresponding small $Y$ deviations of the global vector,
     while the $X$ component is almost linearly decreasing to zero.
\end{itemize}

Notice that, according to this modelling scheme, the path integration
system in Eqs.~(\ref{e09}) and (\ref{e10}) is supposed to work
constantly in the arthropod's neural system during foraging,
reorienting, and homing, except when the animal is seriously perturbed
and not any more able to `measure' and `control' its forward motion
and directional turning. If this happens, the animal is assumed to
instantaneously halt the path integration system and keep the actual
value of the egocentric global vector and the orientational angle
(with respect to an allothetic visual cue) stored until it can proceed
in an unperturbed way.

Let us emphasise, that the presented {\em internal} dynamics of the
global vector $\bm{G}=(X,Y)$ determined by the simple linear system
differential equations (\ref{e09}) and (\ref{e10}), could equivalently
be described in polar coordinates $\bm{G} = r \, (\cos \delta, \sin
\delta)$ using the more complicated nonlinear differential equations
(\ref{e07}) and (\ref{e08}), including the `rotation stop condition'
$\delta = 0$ and the counter-steering term $\omega_0 = - c \, r \cos
\delta$ or a stronger variant like  $\omega_0 = - \tilde{c} \,
\delta$. However, there is an important difference in modelling the
`nest stop condition': In polar coordinates, the obvious termination
criterion would be chosen as $r = 0$ meaning that the global vector
$\bm{G}$ becomes exactly zero. It remains to be proven, which
counter-steering rule could guarantee that this condition is
attainable for stochastically perturbed random paths.

In contrast, the proposed termination criterion $X = 0$ in cartesian
coordinates would, for randomly perturbed return paths, generically
result in a non-vanishing small $Y$ value, then representing the
lateral distance of the arthropod to the nest. Thus, depending on this
value and on the current orientational angle of the animal, the
realized `stop position' can fluctuate around the true nest position,
even in the so far considered case of {\em precise path
  integration}. The size of this random error increases with the
length of the home vector, i.e.\ the distance between food and
nest. This corresponds to experimental observations (e.g.\ for desert
ants {\em C.~fortis} T.~Merkle, unpublished data) which furthermore
show that the length of the outbound path also contributes to such a
positional error. Therefore, other errors in path integration, being
accumulated  along the path, have also to be considered, which is the
topic of the following Section.

\section{Systematic errors}
\label{errorsection}

It has been shown that 
many arthropods \citep[e.g.][]{bis_57,goe_58,hof_85b,weh_86,mul_88}
but also mammals \citep[e.g.][]{seg_93,eti_96,seg_98} exhibit errors
in determining the exact homing direction. In general, we have to
distinguish between random errors and systematic errors during path
integration. There is evidence that random errors, in addition to the
home vector steering error mentioned above, can originate from
inaccurate measurements of angles or distances, whereas systematic
errors probably arise at the neural level of the organism
\citep{ben_90,seg_98}. Orientation is less error prone if allothetic
reference frames are available, as polarised skylight for
arthropods \citep[e.g.][]{weh_98,weh_03a}, but a more difficult task
if not, as for mammals \citep{eti_04}. Systematic errors play an
important role, as the classical two--leg experiments (L-shaped
angular turning tests) have shown in both mammals
\citep[e.g.][]{mau_95,eti_96} and arthropods
\citep[e.g.][]{mul_88,BIS99}. Apart from  mistakes that concern
directional aberrations, there occur also errors 
by underestimation of distances
\citep{som_04}.

From an evolutionary point of view the presence of systematic homing
errors is interesting and has not been explained to date. It may have
an advantage that a homing animal typically assumes a shorter path and
ends up {\em in front of} its nest. It would then avoid an overshoot
and find familiar features that it has just passed on the outgoing
path which might help to reach the nest's entrance.

In this Section we implement two types of systematic errors into our
model. One concerns the estimation of nest distance following
\cite{som_04}, the other exhibit different variations in processing
the turns during path integration. All of them predict systematic
deviations from correct homeward courses and are based on feasible
neural assumptions, or reproduce behaviours that have been observed
during experiments. They may serve as a basis for ongoing and future
studies of systematic deviations (T.~Merkle, unpublished data).

\subsection{Underestimation of turning angles}

\cite{mul_88} trained desert ants to run through two channels of 10m
and 5m length and varied the connecting angle between them in several
steps from $0$ to $180^\circ$, see Figs.~\ref{fig04} and
\ref{fig05}. The ants miscalculated their covered outbound route and,
after leaving the second channel's end, turned about an angle which
was {\em larger} than the correct one leading home. The authors
reproduced this error very well by a simple formula, which accumulates
systematic miscalculations in path integration whenever the animal
walks different from the direct inbound and outbound directions. In
Fig.~\ref{fig05} the angular aberration function $\varepsilon$ is
shown, computed according to the approximative path integration model
by \cite{mul_88}. In general, this function fits quite well
observations in other arthropods and mammals
\citep{seg_93,BIS99}. Here we show that other error models can also
reproduce these data. For evaluation and fitting of the corresponding
error functions (see plots in Fig.~\ref{fig05}) we use the
advantages of our egocentric path integration system.

As a first error mechanism we consider a {\em systematic
underestimation} of  body axis rotation. In principle, that error
could occur during the estimation of $\omega$, i.e.\ by simply
perceiving a value lower than the actual value, or on the neural
level. The high accuracy concerning the ability of desert arthropods
measuring rotations makes it very likely that this error may be
created on the neural level. Therefore, we assume the animal perceives
the correct value $\omega_{\rm real}$, but uses a different value
$\omega_{\rm proc}$ for processing the path integration according to
the differential equations (\ref{e09}) and (\ref{e10}).

In a first choice the underestimation is taken to be a linear function
of the real value,
\begin{equation}
\omega_{\rm proc} = \lambda \; \omega_{\rm real}
\label{1a_underest}
\end{equation}
with a factor $\lambda \! < \! 1$ (error {\sf LU} in Table~\ref{table2}). In
a second variant, the fully saturated underestimation, $\omega_{\rm
  real}$ is processed correctly for small values but saturates towards
a certain maximal turning rate $\omega_{\rm c}$ (error {\sf NLUs} in
Table~\ref{table2}),
\begin{equation}
\omega_{\rm proc} = \frac{\omega_{\rm c}}
{\omega_{\rm c} + |\omega_{\rm real}|} \; \omega_{\rm real}
\label{1b_underest}
\end{equation}
A linear combination of both is given by the error {\sf NLU} in
Table~\ref{table2},
\begin{equation}
\omega_{\rm proc} = \left( \lambda + \frac{1 - \lambda}{\omega_{\rm c} +
  |\omega_{\rm real}|} \; \omega_{\rm c}\right) \; \omega_{\rm real}
\label{1c_underest}
\end{equation}
which again processes small values correctly.

For a related choice of errors we assume a temporal delay $\tau_{\rm
  del}$ in processing the information of $\omega_{\rm real}(t)$; 
the same underestimation could, in principle, be assumed also 
for $v_{\rm real}(t)$, but here variation on
natural outbound paths is rather low (T.~Merkle, personal observations
  on desert ants \emph{Cataglyphis fortis}). Phenomenologically such a
delay is implemented by a linear ordinary differential equation
representing a first order filtering process, namely
\begin{equation}
\frac{d \omega_{\rm proc}}{dt} = \frac{\omega_{\rm real} - \omega_{\rm
    proc}}{\tau_{\rm del}}
\label{1d_delay}
\end{equation}
such that $\omega_{\rm proc}$ is {\em smeared out} on a scale of
$\tau_{\rm del}$ as compared to $\omega_{\rm real}$ (error {\sf PD} in
Table~\ref{table2}).

\subsection{Underestimation of distance to the nest}

The error due to distance underestimation, which we consider here, has
previously been referred to as {\em leaky integrator} by
\cite{som_04}. This idea can be implemented into our egocentric cartesian
path integration model in a straightforward way: with
a constant rate the integrated global vector ``leaks'' or decays from
the memory. Thus, the two--dimensional model Eqs.~(\ref{e09}) and
(\ref{e10}) are varied by simply adding a proportional decay term
in each equation (error {\sf LI} in Table 2)
\begin{eqnarray}
\dot X & = & - v + \omega_{\rm real} Y - \frac{X}{\tau_{\rm L}}
\label{leak_int1} \\
\dot Y & = & - \omega_{\rm real} X - \frac{Y}{\tau_{\rm L}}
\label{leak_int2}
\end{eqnarray}
with mean decay time $\tau_{\rm L}$. Also in egocentric polar
coordinates, c.f.\ Eqs.~(\ref{e07}) and (\ref{e08}), the leaky
integrator is easily expressed by a proportional decay of radial distance
$r$, see Table~\ref{table2}.

In the case of a one--dimensional path, e.g.\ always walking along $x$
without any turns ($\omega_{\rm real} \equiv  0$),
Eqs.~(\ref{leak_int1}) and (\ref{leak_int2}) lead to an exponential underestimation ($x_{\rm
  ue}$) of the actual walking distances ($x$) as
\begin{equation}
x_{\rm ue} = \xi_{\rm L} \left( 1 - \exp(-x/\xi_{\rm L}) \right).
\label{sat_dist}
\end{equation}
The estimated distance $x_{\rm ue}$ saturates at a length $\xi_{\rm
  L} = v \tau_{\rm L}$ in the limit of long walking distances $x$, but
for short paths $x \ll \xi_{\rm L}$ the error is small and $x \approx
x_{\rm ue}$.

This is precisely the best fit to the experiments performed by
\cite{som_04}. The authors trained desert ants to walk through linear
channels to a feeder. Afterwards the ants were captured at the feeder
and released in a linear test channel. They headed off in homeward
direction and performed a back and forth search around their assumed
nest position. By extracting $x_{\rm ue}$ from the search behaviour,
Sommer and Wehner found the relation of Eq.~(\ref{sat_dist}).

In a truly two--dimensional path the leaky integrator of
(\ref{leak_int1}) and (\ref{leak_int2}) may also lead to an {\em
  angular deviation} of the search path from the true homeward 
direction. Earlier sections of the outgoing path have decayed in the
memory more than later ones. If the animal has turned in--between,
this will result in a {\em different} misestimation of related
directions and, consequently, in a homing angle misestimation. In
Fig.~\ref{fig04} this is explained for the classical two--leg
experiment of \cite{mul_88}.

There is, however, a quantitative mismatch between the fit of the LI
equations (\ref{leak_int1}) and (\ref{leak_int2}) to the experiments
of \cite{mul_88} and to those of \cite{som_04}. In the latter case one
obtains $\xi_{\rm L} \approx 90$m which is substantially different
from the value of $18$m of the fit to M\"uller's and Wehner's two--leg
experiments. At present we conclude that most likely some part of the
error occurs during the turn. To fully answer this contradiction, one
would have to take into account more details of the ants' walks, such
as, e.g., walking speed or waiting times.

\begin{table}[!h]
\begin{tabular}[c]{l|l}
\hline
Error & Differential Equations\\
\hline
\hline
{\bf Turning rate underestimation} & \\
\vspace{.1cm}
\begin{tabular}[c]{ll}
{\sf LU} & linear underestimation of $\omega$ \\
&  $ \omega_{\rm proc} = \lambda \; \omega_{\rm real}, \; \; \; \; 0 <
\lambda < 1 $ \\
{\sf NLUs} & fully saturated underestimation \\
& $ \omega_{\rm proc} = \displaystyle\frac{\omega_{\rm c} \omega_{\rm
    real}}{\omega_{\rm c} + | \omega_{\rm real} |} \mbox{ with }
\omega_{\rm c} > 0$\\
{\sf NLU} & partially saturated underest. \\
& $\omega_{\rm proc} = \omega_{\rm  real} \left( \lambda + 
\displaystyle\frac{1 -
  \lambda}{\omega_{\rm c} + | \omega_{\rm real} |} \; \omega_{\rm c}
\right)$ \\
{\sf PD} & processing delay \\
& $\dot \omega_{\rm proc} = \displaystyle\frac{\omega_{\rm real} - \omega_{\rm
    proc}}{\tau_{\rm del}}$
\end{tabular}
&
\begin{minipage}[c]{5cm}
\begin{eqnarray}
\dot X & = & - v + \omega_{\rm proc} \; Y \nonumber \\
\dot Y & = & - \omega_{\rm proc} \; X \nonumber
\end{eqnarray}
\end{minipage}
\\
\hline
{\bf Nest distance underestimation} & \\
\vspace{0.1cm}
\begin{tabular}[b]{ll}
{\sf LI} & Leaky integrator \\
& \\
& egocentric cartesian \\
& \\
& \\
& egocentric polar \\
&
\end{tabular}
&
\begin{minipage}[b]{5.5cm}
\begin{eqnarray}
\dot X & = & - v + \omega_{\rm real} \; Y - X/\tau_{\rm L}
\nonumber \\
\dot Y & = & - \omega_{\rm real} \; X - Y/\tau_{\rm L}
\nonumber
\end{eqnarray}
\begin{eqnarray}
\dot r & = & - v  \cos \delta - r/\tau_{\rm L}
\nonumber \\
\dot \delta & = & v \sin \delta / r - \omega_{\rm real} \nonumber
\end{eqnarray}
\end{minipage}
\\
\hline
\end{tabular}
\caption{Error types in path integration and their respective
  formulae.}
\label{table2}
\end{table}

Up to now we have proposed different elementary error
mechanisms. Based on current knowledge of sensoric and neural
processes it is not possible to prove or refuse their validity. They
nevertheless lead to non--trivial consequences which are not
immediately visible but can be seen in simulated realizations of outbound routes
together with the path integration procedure.

\subsection{Resulting deviations}

\begin{figure}[!ht]
\begin{center}
\includegraphics*[width = 10cm]{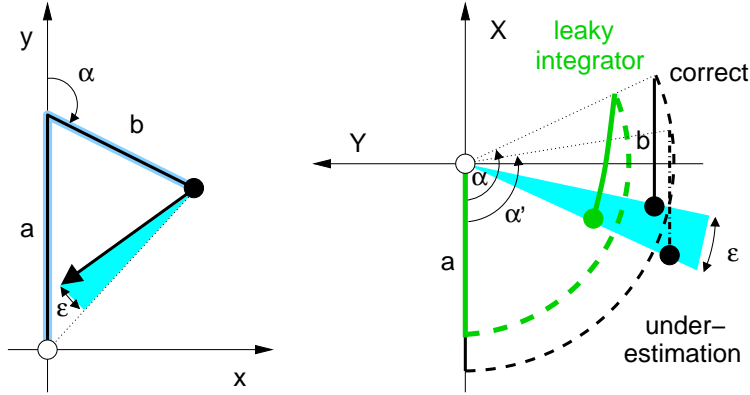}
\end{center}
\caption{Angular deviation $\varepsilon$ in the two--leg experiment
  with channels of length $a$ and $b$, respectively, and clockwise
  connecting angle $\alpha$. (Left side): Experimental situation as
  observed in geocentric $(x,y)$--coordinates. (Right side):
  Representation of the global vector in internal
  $(X,Y)$--coordinates; correct representation indicated by thick
  black line, dashed during counterclockwise turn about angle
  $\alpha$. {\em Angle underestimation} leads to turn by $\alpha' <
  \alpha$ and angular aberration $\varepsilon$ after leaving the
  second channel (black dot--dashed line). {\em Leaky integrator} is
  shown in grey, also leading to angular deviation. For more details
  see text.}
\label{fig04}
\end{figure}

First, we document the outcome of a simulated experiment as in
\cite{mul_88}. The left side of Fig.~\ref{fig04} shows a sketch of
the two--leg experimental setup which in \cite{mul_88} had lengths $a
= 10$m and $b = 5$m. The ant starts at the nest (open circle), turns
after distance $a$ by an angle $0 \le \alpha \le \pi$ to the right,
leaves the channel after another walked distance $b$ (at the black
circle), thereby overcompensating its turn to the correct home
direction by an angular deviation $\varepsilon$.

The right side of Fig.~\ref{fig04} represents this path in the
internal $(X,Y)$--coordinates. First consider a correct processing
without any systematic error: $X$ decreases to $-a$ (thick black
line) and $Y = 0$, then the animal turns by an angle $\alpha$, such
that now $X = - a \cos \alpha$ and $Y = - a \sin \alpha$ (black dashed
line), finally $X$ decreases further to $X = - a \cos \alpha - b$,
whereas $Y$ remains constant (black solid line with black circle).
{\em Angle underestimation} would result in a turn by $\alpha' <
\alpha$ such that in the end $X = - a \cos \alpha' - b$ (dot--dashed
line with black circle) which lies off the true direction to the nest
by an error angle $\varepsilon$. A similar result is obtained by the 
{\em leaky integrator} (fat grey lines): First $X$ decreases from 0 to
$- \xi_{\rm L} \left( 1 - \exp (-a/\xi_{\rm L}) \right)$ and $Y$ remains
0 (thick grey line), then $X$ and $Y$ are turned by an angle $\alpha$
(dashed grey line). The turn occurs so fast that ``leakage'' can be
neglected ($\tau_{\rm turn} \approx r_{\rm turn}/v \ll \tau_{\rm
  L}$, see also below). During the final decrease of $X$, both the
values of $X$ and $Y$ ``leak'' such that finally
\begin{eqnarray}
X_{\rm L} & = & - \xi_{\rm L} \; \left[e^{-b/\xi_{\rm L}} \left( 1 -
  e^{-a/\xi_{\rm L}} \right) \cos \alpha  + \left( 1 - e^{-b/\xi_{\rm
  L}} \right) \right] \\
Y_{\rm L} & = & -  \xi_{\rm L} \; e^{-b/\xi_{\rm L}} \left( 1 -
  e^{-a/\xi_{\rm L}} \right) \sin \alpha
\end{eqnarray}
which is indicated by the filled grey circle, again resulting in an
angular deviation $\varepsilon$ (in the sketch, for simplicity, the
same as for angle underestimation).

\begin{figure}[!ht]
\begin{center}
\includegraphics*[width = 9cm]{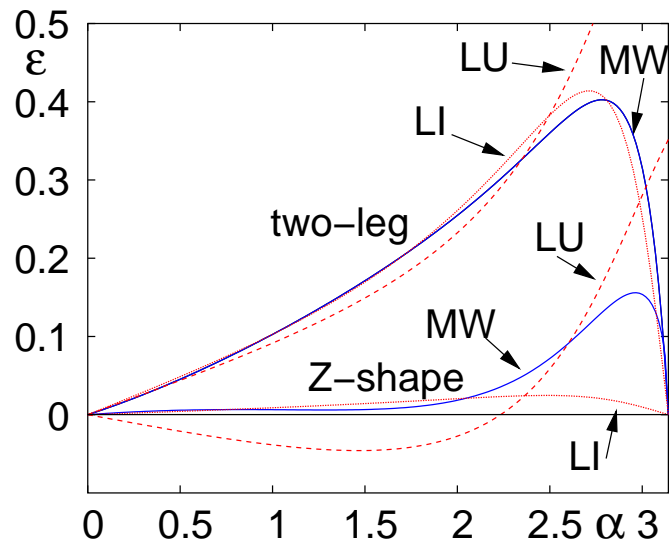}
\end{center}
\caption{Error angles $\varepsilon$ as a function of the
  intermediate turning angle in the two--leg experiment (three upper
  curves) and a Z--shaped channel with three parts of 5~m length each
  (three lower curves) in radian units. Solid lines ({\sf MW}):
  Deviation following \cite{mul_88}; dashed ({\sf LU}): linear
  underestimation of turning rate with $\lambda = 0.87$; dotted ({\sf
  LI}): leaky integrator with $\tau_{\rm L} = 90$ s (resp.\ $\xi_{\rm
  L} = 18$ m). Linear underestimation cannot account for correct path
  integration under full turns ($\alpha = \pi$) but does well for $0
  \le \alpha \le (5/6) \, \pi$. In the Z--shaped channel errors are
  smaller than for single turn and experimentally visible (if at all)
  only for angles around 150$^\circ$ ($= 5 \pi/6$) with errors {\sf
  MW} and {\sf LU}.
}
\label{fig05}
\end{figure}

The three upper curves of Fig.~\ref{fig05} show theoretical
predictions for the angle error $\varepsilon$ in the two--leg
experiment of M\"uller and Wehner: the error according to the formula
of \cite{mul_88} calculated numerically as a function of the angle $0
\le \alpha \le \pi$ between the outgoing channels in radian units
(solid line); the best fit of turning underestimation (dashed line),
i.e.\ $\lambda = 0.87$ in Eq.~(\ref{1a_underest}), and for the leaky
integrator (dash--dotted line), $\tau_{\rm L} = 90$ s in
Eqs.~(\ref{leak_int1}) and (\ref{leak_int2}). The same model errors
are applied to a Z--shaped channel and shown in the three lower curves
with the same coding (solid, dashed, dotted). Note that errors are
smaller, but deviations cancel only partially.

Nonlinear underestimation of the angular turning rate,
Eqs.~(\ref{1b_underest}) and (\ref{1c_underest}), does not show any
different behaviour from (\ref{1a_underest}), because we can assume a
{\em constant} turning rate $\omega_{\rm turn} = v/r_{\rm turn} = 4$
s$^{-1}$ given by the ratio of the walking speed $v = 0.2$ m/s and
radius of the turn in the channel $r_{\rm turn} = 0.05$ m, which is
half the wall to wall distance, because in experiments deserts ants
tend to keep equal distance to both channel walls (T.~Merkle, personal
observation). However, nonlinear underestimation will lead to
different results for arbitrarily curved paths as we will see next.

\begin{figure}[!ht]
\begin{center}
\includegraphics*[width = 9cm]{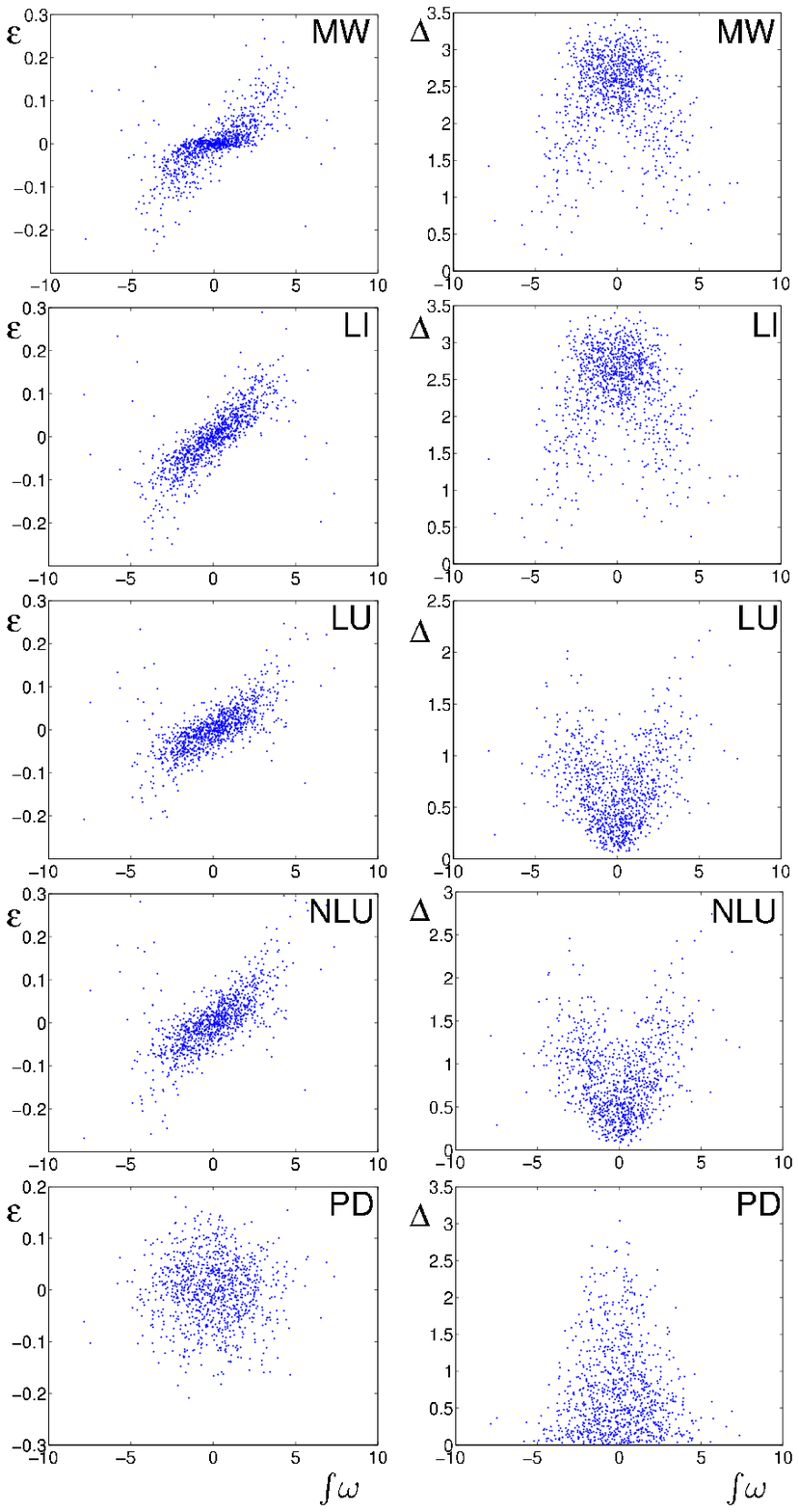}
\end{center}
\caption{Homing error for different path integration error types as
  function of integrated curvature $\int \omega(t) \, dt = \phi_{\rm
  end} - \phi_0$ (in radians, $2 \pi$ for full turn) of 1000 simulated
  random outbound runs of length 20 m each. (Left columns): angular
  deviation $\varepsilon$ in radians; (right columns): euclidean
  distance $\Delta$ from nest in metres. (Top row, {\sf MW}): error of
  \cite{mul_88}; (2nd row, {\sf LI}): leaky integrator with $\tau_{\rm
  L} = 300$ s; (3rd row, {\sf LU}): linear underestimation of
  $\omega(t)$ with $\lambda = 0.87$; (4th row, {\sf NLU}): nonlinear
  underestimation; (bottom row, {\sf PD}): perception delay with
  $\tau_{\rm del} = 0.3$ s exhibits no systematic dependence on
  curvature.}
\label{fig08}
\end{figure}

To investigate how well the different types of systematic errors fit
random outbound paths we simulated runs, as they might be performed by
an untrained ant searching for food without any knowledge on food
sources (see Fig.~\ref{fig06}, upper panel). In particular, we
considered a fluctuating turning rate $\omega(t)$ with a persistence
time $T_\omega$ as in Eq.~(\ref{e12}) of the model in
Section~\ref{modelexcurs}. For 1000 such runs we extracted
correlations between characteristic indicators of the path, such as
its integrated curvature $\phi_{\rm end} - \phi_0 = \int \omega(t) \,
dt$, and the two most direct measures for homing deviation: the {\em
  angular misestimation} $\varepsilon$ between calculated and correct
homeward course, and the {\em euclidean distance} $\Delta$ between
supposed and real nest positions.

The results are shown in Fig.~\ref{fig08}. There is a clear
correlation between curvature $\int \omega$ and the directional
mismatch of the homing vector, for all error mechanisms except for
perception delay (see left panels in column). In particular, all
mechanisms tend to {\em overcompensate} turns effectuated during the
outbound path, as there is a {\em positive correlation} between $\int
\omega$ and the deviation angle. Remind that all predict
overcompensation for the two--leg experiment of M\"uller and Wehner as
well.

There is a striking difference between the leaky path integrator 
and the approximative integration formula of M\"uller and Wehner on
one side, and turning rate underestimation on the other side: The
first two predict a larger euclidean distance from the nest for paths
where left and right turns compensate ($\int \omega \approx 0$) and
come closer to the nest when there is a substantial net turn,
resulting in $\wedge$--shapes in the right hand panels of rows 1 and 2
in Fig.~\ref{fig08}. On the other hand, turning rate underestimation
predicts smaller distance to the nest for compensated turns, and
larger euclidean mismatch for paths with higher turns, leading to
$\vee$--shapes of lines 3 and 4 in the right hand panels. Experiments
which cover both the {\em full return path} and the {\em systematic
  search}, additionally to the initial direction analysed by
\cite{mul_88} may be able to decide the type of homing error mechanism
in desert ants (T.~Merkle, in preparation).

\begin{figure}[!ht]
\begin{center}
\includegraphics*[width = 5cm]{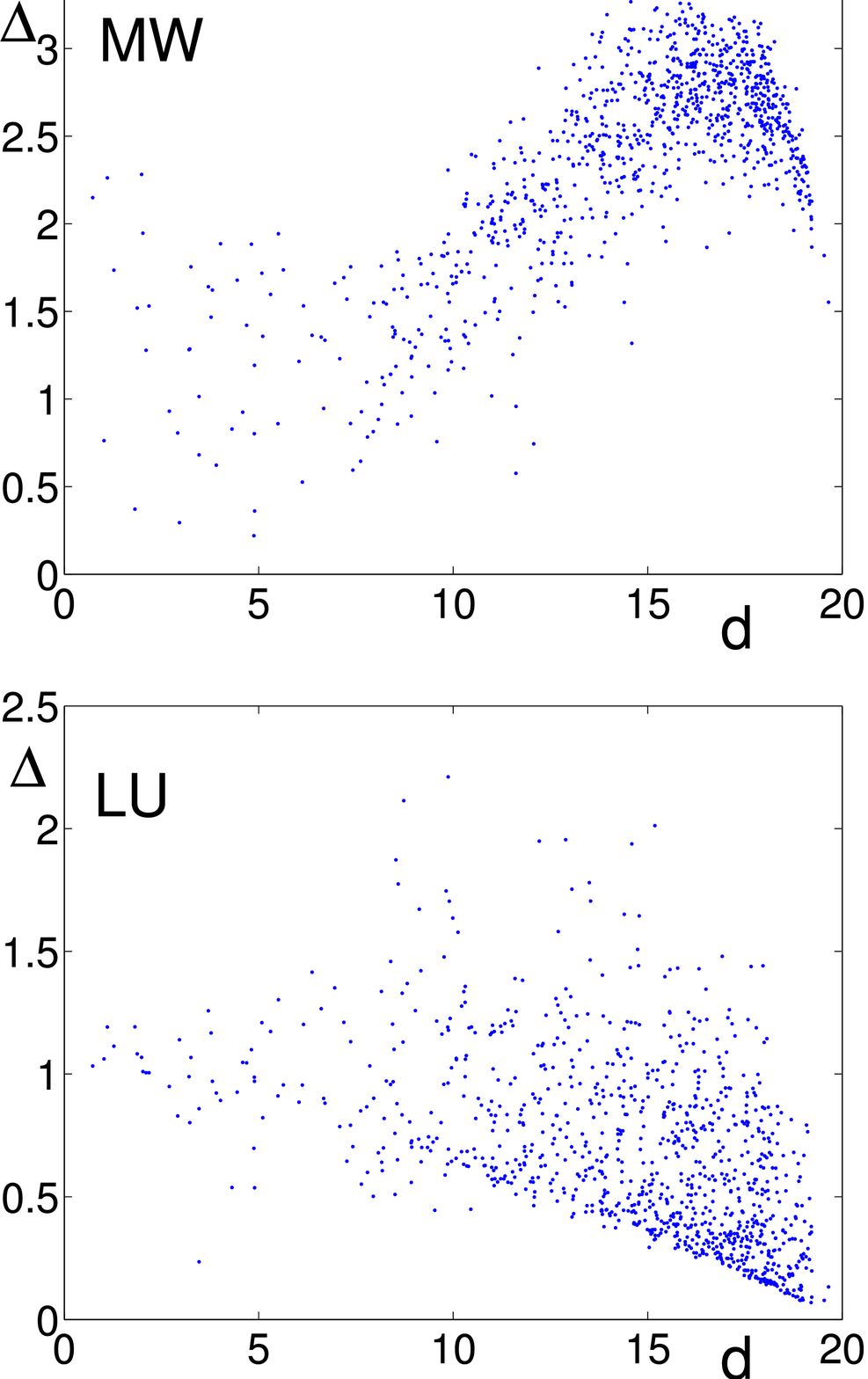}
\end{center}
\caption{Homing error in euclidean distance $\Delta$ between supposed
  and  real nest position as function of distance between starting
  and end point of foraging path, i.e.\ distance of feeding site from
  nest, $0 \le d \equiv |\bm{P}(t_1) - \bm{P}x(0)| \le 20$. Simulated
  paths had arch length 20~m, so $d = 20$~m means a perfectly straight
  path. (Top): The {\em leaky integrator} predicts increasing $\Delta$
  with $d$, whereas  according to \cite{mul_88} $\Delta$ has a maximum
  for intermediate $d$ (middle). (Bottom): {\em Angle underestimation}
  leads to an opposite relation, $\Delta$ decreasing with $d$. Same
  parameters as in Fig.~\ref{fig08}.}
\label{fig08a}
\end{figure}

Fig. \ref{fig08a} shows the homing errors produced by different error
types as functions of the distances $d$ between start and end points
of foraging trips that had the same overall path lengths. Thus, the
values of $d$ indicate the sinuousity of the different paths: straight
paths have large, winded paths small $d$. Roughly speaking the error
of the {\em leaky integrator} increases with $d$ and becomes maximal
for perfectly straight paths. The error postulated by \cite{mul_88}
also increases with $d$ over a wide range, but {\em decreases} for
very large values for almost straight paths. {\em Angle
  underestimation} yields an opposite picture, the deviation decreases
over the entire range of $d$, although large fluctuations may
obfuscate measurements. Clearly these findings have to be further
developed in comparison to real experiments, but they indicate how
field work can enable an observer to differentiate between various
error types.

\begin{figure}[!ht]
\begin{center}
\includegraphics*[width = 10cm]{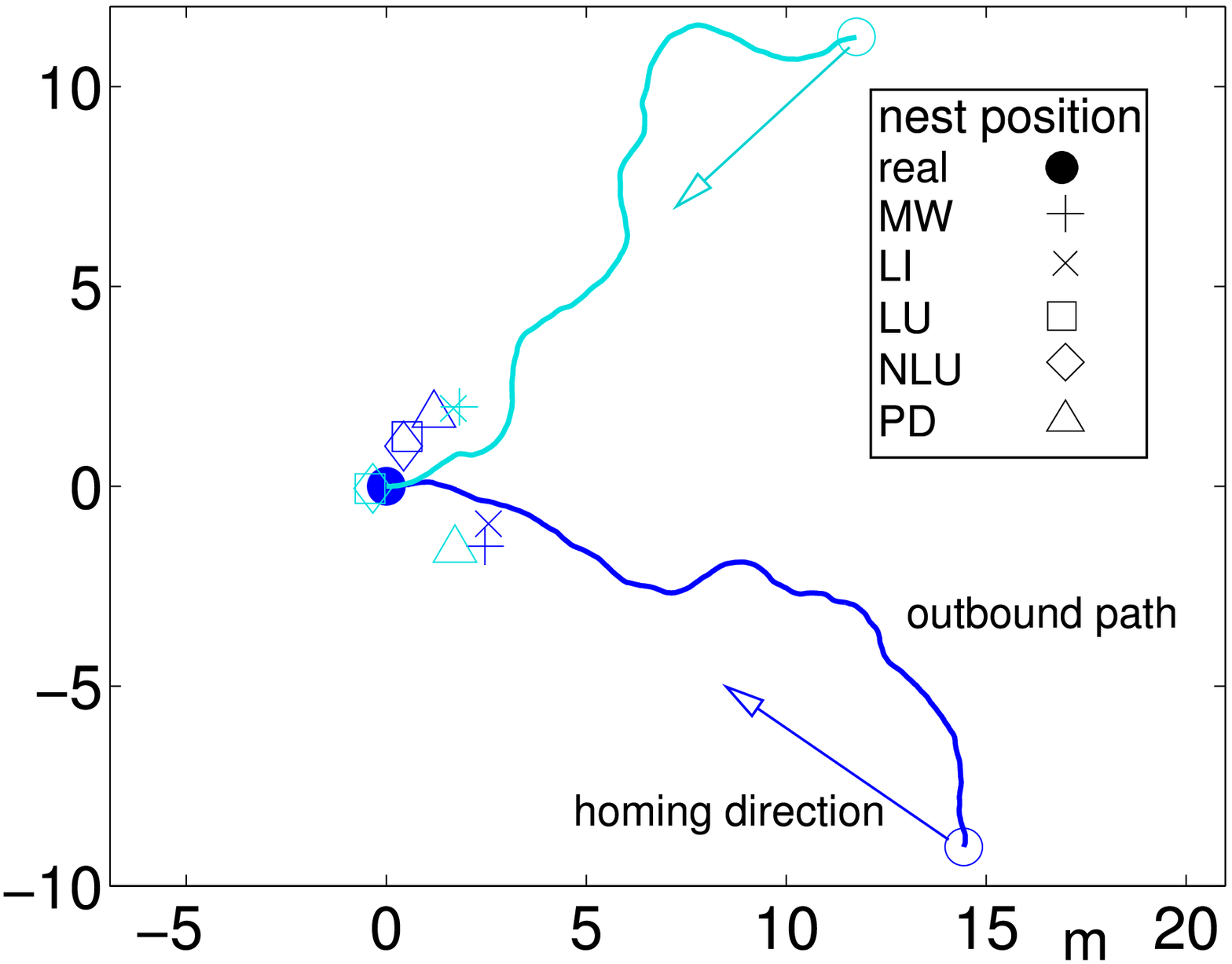}
\includegraphics*[width = 10cm]{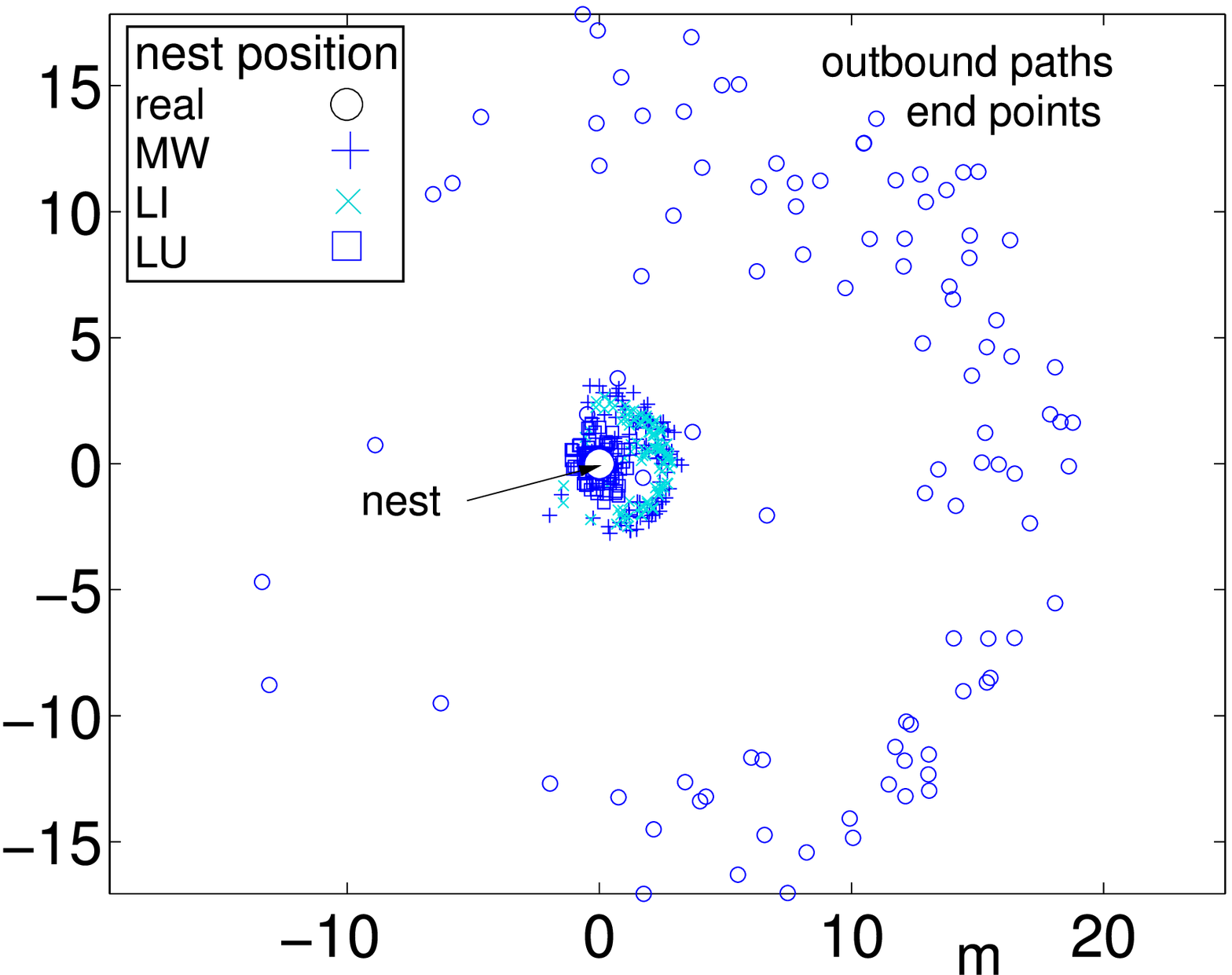}
\end{center}
\caption{Supposed  nest locations for different error types. (Upper
  panel): Two simulated random outbound paths both starting in
  direction $\phi_0 = 0$ and having arch length 20 m (black and grey)
  with supposed nest locations according to M\"uller and Wehner ($+$),
  leaky integrator with $\tau_{\rm L} = 300$ s ($\times$),
  underestimation of $\omega$ with $\lambda = 0.87$ and $\omega_{\rm
  c} = 0.2$ s$^{-1}$ (linear: $\Box$, nonlin.: $\Diamond$) and
  processing delay of $\omega$ with $\tau_{\rm del} = 0.3$ s
  ($\triangle$). Note that ($+$) and ($\times$) are relatively
  close. (Lower panel): End points of 100 simulated outbound paths
  ($\circ$) with supposed nest positions after M\"uller and Wehner
  ($+$), leaky integrator (grey $\times$), and
  $\omega$--underestimation ($\Box$). Real nest marked by filled white
  circle. Note that ($+$) and ($\times$) coincide well, in front of
  real nest. ($\Box$) are grouped closely around the nest. Units are
  in metres.}
\label{fig06}
\end{figure}
A difference is also visible in the predictions for the supposed nest
positions, as presented in Fig.~\ref{fig06}. In its upper panel it
shows the endpoints of two random outbound runs of length 20 m (one in
black, the other one in light grey), together with the respective
supposed nest locations under different error mechanisms: the formula
of M\"uller and Wehner (MW, marked by $+$), linear (LU, $\Box$) and
nonlinear (NLU, $\Diamond$) underestimation with $\lambda = 0.87$ and
$\omega_{\rm c} = 0.2$ s$^{-1}$ and turning perception delay with
$\tau_{\rm del} = 0.3$ s (PD, $\triangle$). The time constant for the
leaky integrator (LI, $\times$), $\tau_{\rm L} = 300$ s, was chosen
such that it best fitted the results of the phenomenological error
formula of \citep{mul_88}. In the lower panel of Fig.~\ref{fig06} the
same is shown for 100 paths (without the paths themselves), where all
runs start in the same initial direction $\phi_0 = 0$, such that the
end points $(\circ)$ lie in a sickle shaped domain to the right. Again
there is a striking coincidence between the leaky integrator and
M\"uller's formula as opposed to the results of turning rate
underestimation. Notice that only the two first error mechanisms lead
to a home vector pointing to a location {\em in front of} the actual
nest (see small sickle--shape domain to the right of nest position).

\section{Discussion}
\label{discussion}

In this work we have presented a very simple model for path
integration using egocentric cartesian coordinates. In contrast to all
previous models, including the egocentric one using polar coordinates
(Section~\ref{prev_egocent}), in this model the arthropod does not
need to perform complicated calculations such as applying
trigonometric or other non--linear functions, but rather updates two
cartesian coordinate values of the relative global vector
$\bm{G}=(X,Y)$ by computing a simple system of linear differential
equations. Moreover, although we assume that neither the actual
relative angle $\delta$ nor the distance $r$ to the nest have to be
calculated or stored at any time, solely by using the internal
$\bm{G}$--vector information the arthropod has the ability to orient
towards the nest position at any time along its path and to hold this
orientation during the home run. Keeping $Y = 0$ serves as an
`internal beacon' for home orientation, where the simple
counter--steering mechanism can be realized as an elementary negative
feedback control of the turning rate by the internal variable $Y$,
until the second internal variable $X$ reaches the desired zero
value. Thus, the path integration values $(X,Y)$ do not only provide a
record of the arthropod's positional movement, but can also be used as
information input for orientation. Moreover, accumulated information
on the whole internal $(X,Y)$-path, as depicted in Fig.~\ref{fig02},
for example, may be used by the arthropod to guide its observed
systematic search for the `true' nest position after failure. This
could be done by estimating the probability of how far away the nest
is located, depending on the probability of the accumulated path
integration error. For previous experiments and theories see
\cite{HOF90}.

In order to prepare the theoretical background for further
investigations, we have restricted this first presentation to (i)
construct our model, (ii) implement various error mechanisms and (iii)
evaluate their predictions for future comparison with experiments by
stochastic simulations. Figs.~\ref{fig08} and \ref{fig08a} suggest
simple checks to accept or exclude one or another error mechanism and
therefore can serve as a guide for future experiments and modelling.

The mechanisms suggested here could also be tested in experiments with
specially designed channels for outbound runs. Different types of
angle misestimation would e.g.\ lead to different systematic
deviations of the home run in a channel where left turns are sharper
than right turns. Differences between the leaky integrator and
angle--misestimation should become apparent in a comparison between
two different two--leg experiments, e.g.\ both with $\alpha = \pi/2$,
but two different values $a_1 \neq a_2$ which both would be
substantially larger than $b$.

Physiological realizations of the integration procedure itself and the
underlying fundamental neural mechanisms are far from being
clarified. Besides, a clarification of the neural processes working in
the brain and the locomotory  control apparatus of desert arthropods
does not seem to be within reach in the near future. We share this
problem with all other existing models of path integration.

On the other hand, anatomic features of neurons, their interactions,
their integration, and their cooperation within networks have been
known for a long time. \cite{har_95} have developed a simple and
efficient neural network for path integration in desert ants. In a
particular form it even incorporates the systematic errors observed by
\cite{mul_88} on a neural level. Based on their model of an
incremental encoding one could easily construct a neural architecture
for path integration in cartesian variables $X$ and $Y$ as in
Eqs.~(\ref{e09}) and (\ref{e10}), where the required estimation of
speeds could most likely be encoded by spike rates of afferent
neurons.

A further, much simpler way would be to represent the internal
cartesian coordinates $(X,Y)$ directly by the deviations of two
non--spiking interneuron activities $N^X$ and $N^Y$ from their basal
activity values $N^X_0$ and $N^Y_0$  supposed to be attained when the
animal is `at home', i.e. for $(X,Y) = (0,0)$. Regulation of these
interneurons as well as their mutual interactions could then be
realized by suitably defined dendritic synapses of a neural net akin
to the scheme presented in Fig.~\ref{fig03}. Again, the obvious
simplicity of this `linear'  control network may favour the egocentric
cartesian path integration model as candidate for a most elementary
neural realization in the arthropod, compared to other, more
complicated models. Although mathematical simplicity is not an ad--hoc
argument to explain natural evolution of biological control systems,
it is tempting to ``grow'', i.e.\ let develop by evolutionary
algorithms, neural networks for the task of orientation and analyse
their mathematical structure post--hoc, as it has been done for robot
motion control \citep{pas_01}.

The ability to find successful feeding sites, also observed for
beetles, for instance \citep[von][]{FRI65}, was particularly
investigated in detail for desert ants
\citep[][]{WEH83b,WEH87,col_99}. It is therefore a natural question to
ask, whether a similar simple rule as that of keeping $Y = 0$ may help
to find a previously known feeding site. Consider all trajectories in
geocentric coordinates which keep the egocentric $Y \equiv 0$ constant:
they are the radii around the nest position. On the way home they all
{\em converge}, and if by random fluctuations the animal switches over
to a trajectory in its neighbourhood it nevertheless is guided towards
the nest by the beacon condition $Y = 0$. But for outbound routes they
{\em diverge}, and random errors are not corrected on their own. It
would even be better to follow a {\em fixed compass direction},
because trajectories of the same direction are parallel to each other,
and randomly accumulated errors will not be enhanced during the
course. In the light of our model it seems natural to suggest that the
relative position of a feeding site is internally stored as another
global vector $\bm{G}_{\rm f} = (X_{\rm f},Y_{\rm f})$ which is
updated simultaneously with $(X,Y)$. Depending on whether the animal
steers towards the feeder or home, either $Y_{\rm f} = 0$ with $X_{\rm
  f} > 0$ is the beacon condition, or $Y = 0$ with $X>0$. To our
knowledge no experiments with obstacles on the way to a trained feeder
have been performed corresponding to those described in \cite{weh_03a}
for homing paths with obstacles (see Fig.~2B therein). If the animals
are able to compensate forced deviations on the path to the feeder in
the same manner as on paths leading home, this would indicate a
similar internal processing for both positions.

Moreover, a better understanding of mechanisms by which global vector
information is combined or substituted with local information 
from landmarks requires efficient mathematical models and
simulations that are able to reproduce experimental data. With our
egocentric path information system for the two relative cartesian
coordinates we have presented a most simple modelling tool that can
help to evaluate and discriminate various hypotheses on orientation,
random or systematic errors, and possible neural representations.

\subsection*{Acknowledgements}

This work was partly supported by Research Group {\em Wissensformate}
of Bonn University and Special Research Program SFB 611 of Deutsche
For\-schungs\-ge\-mein\-schaft. We thank G.~Hoffmann for helpful
comments on the manu\-script, the anonymous referees for many detailed
suggestions to improve the manuscript, and R.~Wehner for introducing
T.M.\ to {\em Cataglyphis} and including him in his research
project on path integration in desert ants.

\bibliographystyle{JTB}

\bibliography{JTB}

\end{document}